\documentclass{article}
\usepackage{amsmath,amssymb,amsfonts}
\usepackage{graphicx}
\usepackage{textcomp}
\usepackage{dsfont}
\usepackage{xcolor}
\definecolor{A01}{RGB}{1, 88, 143}
\definecolor{A02}{RGB}{40, 106, 153}
\definecolor{A03}{RGB}{77, 124, 163}
\definecolor{A04}{RGB}{115, 142, 173}
\definecolor{A05}{RGB}{153, 160, 182}
\definecolor{A06}{RGB}{190, 178, 192}
\definecolor{A07}{RGB}{228, 196, 201}
\definecolor{B01}{RGB}{1, 71, 119}
\definecolor{B02}{RGB}{38, 85, 128}
\definecolor{B03}{RGB}{74, 99, 136}
\definecolor{B04}{RGB}{111, 114, 144}
\definecolor{B05}{RGB}{148, 128, 152}
\definecolor{B06}{RGB}{185, 142, 160}
\definecolor{B07}{RGB}{222, 156, 168}
\definecolor{C01}{RGB}{1, 53, 96}
\definecolor{C02}{RGB}{37, 64, 103}
\definecolor{C03}{RGB}{73, 75, 110}
\definecolor{C06}{RGB}{180, 108, 129}
\definecolor{C07}{RGB}{216, 118, 135}
\definecolor{D01}{RGB}{69, 139, 116}
\definecolor{D07}{RGB}{127, 255, 212}
\definecolor{E01}{RGB}{1, 18, 49}
\definecolor{E02}{RGB}{35, 22, 52}
\definecolor{E03}{RGB}{68, 25, 55}
\definecolor{E04}{RGB}{102, 29, 59}
\definecolor{E05}{RGB}{136, 33, 62}
\definecolor{E06}{RGB}{169, 36, 65}
\definecolor{E07}{RGB}{202, 40, 68}
\definecolor{F01}{RGB}{1, 0, 25}
\definecolor{F02}{RGB}{34, 0, 27}
\definecolor{F03}{RGB}{66, 0, 28}
\definecolor{F04}{RGB}{99, 0, 30}
\definecolor{F05}{RGB}{131, 0, 32}
\definecolor{F06}{RGB}{163, 0, 33}
\definecolor{F07}{RGB}{196, 0, 35}
\usepackage{diagbox}
\usepackage{subfig}
\usepackage{caption}
\usepackage{tikz}
\usetikzlibrary{shapes,arrows}
\usetikzlibrary{calc}
\usepackage{todonotes}
\usepackage{nomencl}
\usepackage{ulem}
\usepackage{algorithm, algpseudocode}

     \PassOptionsToPackage{numbers, compress}{natbib}
     \usepackage[preprint]{neurips_2024}

\usepackage[utf8]{inputenc} % allow utf-8 input
\usepackage[T1]{fontenc}    % use 8-bit T1 fonts
\usepackage{hyperref}       % hyperlinks
\usepackage{url}            % simple URL typesetting
\usepackage{booktabs}       % professional-quality tables
\usepackage{amsfonts}       % blackboard math symbols
\usepackage{nicefrac}       % compact symbols for 1/2, etc.
\usepackage{microtype}      % microtypography
\usepackage{xcolor}         % colors

\title{Data is missing again -- Reconstruction of power generation data using $k$-Nearest Neighbors and spectral graph theory}

\author{%
  Amandine Pierrot\thanks{Amandine Pierrot is now with the Department of Mathematical Sciences, University of Bath, UK.\\ 
  Email: \texttt{amcp23@bath.ac.uk}}\\
  Department of Wind and Energy Systems\\
  Technical University of Denmark\\
  Denmark\\
  \texttt{} \\
  \And
  Pierre Pinson\thanks{Pierre Pinson is also with Halfspace, DK, the Department of Technology, Management and Economics, Technical University of Denmark, DK, and CoRE, Aarhus University, DK.\\ 
  Email: \texttt{p.pinson@imperial.ac.uk}}\\
  Dyson School of Design Engineering\\
  Imperial College London\\
  United Kingdom\\
}

\makenomenclature

\begin{document}

\bibliographystyle{jasa3}

\maketitle

\begin{abstract}
  The risk of missing data and subsequent incomplete data records at wind farms increases with the number of turbines and sensors. We propose here an imputation method that blends data-driven concepts with expert knowledge, by using the geometry of the wind farm in order to provide better estimates when performing Nearest Neighbor imputation. Our method relies on learning Laplacian eigenmaps out of the graph of the wind farm through spectral graph theory. These learned representations can be based on the wind farm layout only, or additionally account for information provided by collected data. The related weighted graph is allowed to change with time and can be tracked in an online fashion. Application to the Westermost Rough offshore wind farm shows significant improvement over approaches that do not account for the wind farm layout information.
\end{abstract}

\section{Introduction}\label{sec:intro}
According to the International Energy Agency, overall wind power generation increased by a record 17\% in 2021. Of the total 830 GW installed, 93\% were still onshore systems, as onshore wind is a developed technology while offshore wind is still at the early stage of expansion. However, offshore reach is expected to increase in the coming years as more countries are developing or planning to develop their first offshore wind farms. From the world's first offshore wind farm, Vindeby in Denmark, which totalled 11 turbines in 1991, the size of offshore wind farms has increased up to more than a hundred wind turbines nowadays, e.g., Hornsea 1 in the United Kingdom that totals 174 wind turbines. While data recorded by wind turbines are of great value for wind farm and system operators, they are subject to information loss from, e.g., power and communication failures, instrumentation issues or human error. Missing data in wind farm time series can impact revenue \cite{Coville2011}, wind energy resource assessment \cite{Salmon2014}, wind farm control \cite{Hosseini2014} or the estimation of power curves \cite{Hu2019}. In particular, they negatively impact forecasting models, which for short-term lead times (from a few minutes to a few hours ahead) are better be statistical models trained on historical data, or online learning methods that require the most recently observed data \cite{Tawn2020}. Because of the increasing number of turbines in offshore wind farms, the issue of missing data gets even more critical. Let $T$ be the total number of records over a wind farm, measured at successive time steps $t=1,\dots,T$ (usually spaced at uniform intervals). Now assume a data point is missing for a wind turbine at time $t$ with probability 0.01, independently from other wind turbines. With a number of wind turbines $N=11$, this would result in about 90\% of the $T$ records being complete, i.e., data points are available for all $N$ wind turbines.  With $N=174$ wind turbines, the proportion of complete records drops to 17\% and the workaround that consists of assuming data completeness and deleting records with missing entries is not sensible anymore \cite{Zhu2022}. 

Alternatives remain for dealing with increasing missing data. One is to develop methods where the assumption of data completeness is not needed anymore. In the context of time series, works exist that make assumptions about the missing data patterns \cite{Dunsmuir1981}, or need not even make any assumptions \cite{Anava2015}, and estimate AR models. Other works develop models that are robust to missing data \cite{Stratigakos2023}. Another alternative is to provide imputations for missing values, i.e., to replace missing data points with plausible values. Classical statistical imputation methods use maximum likelihood estimators that correspond to a specific underlying model. A very popular approach for dealing with missing data in time series is the EM algorithm \cite{Dempster1977}, which relies on two steps: at the E-step, missing values are filled in with their conditional expectation given the observed data and the current estimate of the model parameters; at the M-step, new estimates of the parameters are computed from the current version of the completed data. This procedure requires assumptions on the distributions of both observed and missing data. A widely used \cite{Rubin1976}, yet controversial \cite{Seaman2013}, nomenclature for missing value mechanisms distinguishes between three cases: MCAR, MAR and MNAR. In MCAR, the probability of a data point being missing is completely independent of any variables in the dataset, while in MAR the probability of being missing depends only on observed values. These first two mechanisms are considered the simple ones, in the sense that they do not make it necessary to model the distribution of the missing values when maximizing the likelihood of the observations. The third mechanism is the harder yet prevalent one, as the probability of a point being missing is dependent on the value it would have taken. This leads to important biases in the remaining data whose distribution is not the true distribution anymore. Missing value imputation is appealing because it makes it possible to first get a completed dataset, and then apply any statistical learning algorithm that relies on the completeness assumption. Yet, there is a wide range of situations where it might be more or less legitimate to use imputation. This has to do not only with the missing value mechanism, but also with the task to be performed on the completed dataset. For supervised learning tasks such as regression (e.g., for forecasting, eventually), theoretical and empirical results outline simple practical recommendations \cite{Josse2020} when using imputation methods. In particular, the same imputation model should be used to train and test on data with missing values. Empirically, better imputation methods seem to reduce the number of samples required to reach good prediction. When the supervised learning algorithm is of the regression kind, almost all imputations lead asymptotically to the optimal prediction with a powerful learner, no matter the missing value mechanism. This result gives theoretical grounding to all impute-then-regress procedures. Yet, a good choice of imputation can reduce the complexity of the regression function to be learned and therefore it is suggested that learning imputation and regression jointly is easier \cite{Morvan2021,Wen2022}. 

In the context of offshore wind farms, we deal with multivariate time series, as we record $N$ data points, one for each wind turbine, at each time step $t$. This opens a new range of methods for missing data imputation, as one can exploit information from another (potentially correlated) sensor, in our case another wind turbine. Recently, several deep learning approaches have been proposed for multivariate time series imputation \cite{Cini2022, Luo2018, Cao2018}. When interested in the average production of a wind farm, it is quite intuitive to work with the average of the individual production values that are available at time $t$. By doing so, one implicitly performs $k$-NN imputation. The $k$-Nearest Neighbors algorithm is a seminal nonparametric method in machine learning \cite{Bible2009, Biau2015}. In a nutshell, it uses the $k$ closest points to a point of interest to make a decision about the latter. When using $k$-NN for imputation purposes, one considers the $k$ nearest neighbors of a missing point to provide an estimate of its value \cite{Troyanskaya2001}. The assumptions associated with this imputation method are very weak: we do not need assume any model generating the data, observed or missing, and only assume similar groups of observations. Moreover, the method applies for all missing data mechanisms. In Section \ref{sec:method}, we make it explicit how to work with a quantity of interest averaged over $n_t$ available records comes down to performing unweighted $n_t$-NN imputation. We propose to improve it by moving from unweighted to weighted $n_t$-NN imputation through Nadaraya-Watson estimators. Each neighbor will now enter the $k$-NN algorithm with a different weight, hopefully the closer the higher. A higher weight for a closer neighbor means we are able to measure how close with an appropriate distance. We show how to use graph spectral theory to compute Laplacian eigenmaps, i.e., new representations of the wind farm as a graph that take into account local and global geometries. We consider the case where we only use the structure of the wind farm when learning its representation, and the case where we also use values of the quantity we wish to perform imputation for. Regarding the latter, we focus on power generation, but other missing quantities could be considered, e.g., wind speed. The method is illustrated on the Westermost Rough offshore wind farm and results for the imputation of power generation missing values are presented in Section \ref{sec:results}.  Finally, we provide some conclusions and perspectives in Section \ref{sec:conclusion}.

\renewcommand\thefootnote{}
\footnotetext{\textbf{Abbreviations: NN, nearest neighbors; AR, autoregressive; EM, expectation-maximization; MCAR, missing completely at random; MAR, missing at random; MNAR, missing not at random; OGD, online gradient descent; RMSE, root mean square error}.}

\section{Imputation using Nearest Neighbors and graphs}\label{sec:method}
In this section, we introduce the different estimators at hand to perform $n_t$-NN imputation when interested in the power generation over a wind farm. They are summarized in Table \ref{tab1}.
\begin{table*}[!h]%
\centering %
\caption{Estimators for the $n_t$-NN imputation of power generation's missing values.\label{tab1}}%
\begin{tabular*}{\textwidth}{@{\extracolsep\fill}lll@{\extracolsep\fill}}
\toprule
\textbf{Estimator} & $k$\textbf{-NN}  & \textbf{Graph} \\
\midrule
Naive & Unweighted & No  \\
Location & Weighted & No \\
Unweighted-graph & Weighted & Yes, unweighted \\
Weighted-graph & Weighted & Yes, weighted \\
\bottomrule
\end{tabular*}
\end{table*}
Weighted $k$-NN imputation is introduced in Section \ref{sec:weightedNN}, unweighted graphs are introduced in Section \ref{sec:unweightedgraphs} and weighted graphs in Section \ref{sec:weightedgraphs}. The naive and location-based methods are to be seen as standard ones, as opposed to the proposed graph-based methods. The naive estimator is an unweighted $k$-NN benchmark, while the location estimator is a weighted $k$-NN benchmark.

\subsection{Weighted Nearest Neighbor imputation}\label{sec:weightedNN}
Let $X_t$ be an average quantity of interest over a wind farm at time $t$. We have 
\begin{equation}
    X_t = \frac{1}{N} \sum_{i=1}^N X_t^{i},
\end{equation}
where $X_t^i$ is the quantity of interest for the $i$-th wind turbine at time $t$ and $N$ is the total number of wind turbines in the wind farm. When some of the $X_t^i$s are missing, assume  that we work instead with the estimate
\begin{equation}\label{eq:available.avg}
    \Hat{X}_t = \frac{1}{n_t} \sum_{j=1}^{n_t} X_t^{(j)},
\end{equation}
where $X_t^{(j)}$ is the quantity of interest for the $(j)$-th wind turbine record available at time $t$, $n_t$ being the number of available wind turbine records. Imputation using the $k$-NN method consists of filling in a missing value using the values from its $k$ nearest neighbors. Unweighted $k$-NN assign the same weight to every neighbor, when weighted $k$-NN assign a higher weight to a closer neighbor. Let us replace each missing value $X_t^{(l)}$ with its unweighted $n_t$-NN estimate $\Hat{X}_t^{(l)}=\frac{1}{n_t}\sum_{j=1}^{n_t}X_t^{(j)}$. We have
\begin{subequations}
    \begin{align}
    \Hat{X}_t &= \frac{1}{N}\left(X_t^{(1)} + \dots + X_t^{(n_t)} + \Hat{X}_t^{(n_t + 1)} + \dots + \Hat{X}_t^{(N)}\right),\\
    &= \frac{1}{N}\sum_{j=1}^{n_t} X_t^{(j)} + \frac{1}{N}\sum_{l=n_t+1}^{N} \Hat{X}_t^{(l)},\\
    &= \frac{1}{N}\sum_{j=1}^{n_t} X_t^{(j)} + \frac{1}{N}\sum_{l=n_t+1}^{N} \frac{1}{n_t}\sum_{j=1}^{n_t}X_t^{(j)},\\
    &= \frac{1}{n_t} \sum_{j=1}^{n_t} X_t^{(j)}.
\end{align}
\end{subequations}
Hence, to work with $\Hat{X}_t$ from \eqref{eq:available.avg} is equivalent to filling in the missing values $(X_t^{(l)})_{l=n_{t + 1},\dots,N}$ using unweighted $n_t$-NN estimates, i.e., every neighbor $X_t^{(j)}$ is assigned the same weight $1/n_t$. However, a quantity of interest at a wind turbine level is likely to be more similar to the same quantity from the actual neighbors of this wind turbine, i.e., the wind turbines that are nearby in the wind farm. Staying in the $k$-NN framework, we can improve our estimates through the number of neighbors $k$, the weights assigned to neighbors, or both. Theoretical results about $k$-NN mostly concern the asymptotic mode, when $n_t$ tends to infinity, which cannot be assumed here as we are limited by the number of wind turbines in the wind farm. It is rather critical to choose $k$ in a finite regime, and it is usually advised to perform cross-validation. This would be cumbersome in our setup as cross-validation would need to be run for each combination of available data points, for each missing data point, and would require enough complete data for each combination. Therefore, we propose to keep $k=n_t$ at each time $t$ and to rather improve the weights of the $n_t$-NN imputation. Learning the distance metric for $k$-NN has been extensively studied and it has been found that metric learning may significantly affect the performance of the method in many applications. We refer the interested reader to reviews of the metric learning literature \cite{Kulis2012} and the $k$-NN method literature \cite{Biau2015}. Because we perform imputation at each time step $t$ considering values from similar sensors at the same time step $t$, the Euclidean distance seems a fair enough metric in our framework. Therefore, we focus instead on a common shortcoming in current nonparametric methods, which is to only consider the distances between the decision point and its neighbors and ignore the geometrical relation between those neighbors. Indeed, before we even get any records from its sensors, a wind farm is a graph with its own geometry that provides a priori information not only on the distance between a wind turbine and its neighbors, but also between these neighbors. 

Moving to weighted $n_t$-NN imputation, we wish to provide each wind turbine $(l)$ that misses a record with a better estimate, by weighting the available records $(j)$ according to their proximity to the wind turbine, while acknowledging the whole structure of the wind farm. In order to do so, we need to be able to assign weights depending on the distance between the wind turbine $(l)$ and the wind turbines $(j)$. We choose to use Nadaraya-Watson estimators \cite{Nadaraya1964, Watson1964}, which assign weights that are proportional to some given similarity kernel $K$. More optimal methods could be used \cite{Anava2016}, which we will discuss later. Let $K$ be a given nonnegative measurable function on $\mathbb{R}$ (the kernel), $h$ be a positive number (the bandwith) depending upon $n_t$ only and $\lVert \textbf{z}^{(j)} - \textbf{z}^{(l)} \rVert$ be the Euclidean distance between the representations of two wind turbines $(j)$ and $(l)$. In case $(l)$ is missing and $(j)$ is available at time $t$, the weight we give to $X_t^{(j)}$ when computing a weighted estimate $\Hat{X}_t^{(l)} = \sum_{j=1}^{n_t} w^{(jl)}X_t^{(j)}$ of $X_t^{(l)}$ is 
\begin{equation}
    w^{(jl)} = \frac{K\left(\frac{\lVert \textbf{z}^{(j)} - \textbf{z}^{(l)} \rVert}{h}\right)}{\sum_{i=1}^{n_t} K\left(\frac{\lVert \textbf{z}^{(i)} - \textbf{z}^{(l)} \rVert}{h}\right)}.
\end{equation}
Let us sort the neighbors of a wind turbine $(l)$ by increasing distance, $\lVert \textbf{z}^{(1)} - \textbf{z}^{(l)} \rVert \leq \lVert \textbf{z}^{(2)} - \textbf{z}^{(l)} \rVert  \leq \dots \leq \lVert \textbf{z}^{(n_t)} - \textbf{z}^{(l)} \rVert$. We choose an adaptive bandwith $h_t=\lVert \textbf{z}^{(n_t)} - \textbf{z}^{(l)} \rVert$, so that the weights adjust depending on $n_t$, i.e., depending on the availability of other records at time $t$. Note that if all distances at play at time $t$ are very similar, the estimator will be very close to the unweighted one, which seems legit. Consider the so-called naive kernel $K(\textbf{u})=\mathds{1}_{\{\lVert \textbf{u} \rVert\leq 1\}}$. With such a choice for $h_t$, to use a naive kernel is to use our former estimate $\frac{1}{n_t}\sum_{j=1}^{n_t}X_t^{(j)}$. Therefore\textcolor{blue}{,} from now on we will refer to this estimate as the Nadaraya-Watson estimator with a naive kernel, i.e., the ``naive'' estimator. For a more general kernel, the weight $w_t^{(jl)}$ depends on the distance $\lVert \textbf{z}^{(j)} - \textbf{z}^{(l)} \rVert$ through the kernel shape. We will consider the 
\begin{itemize}
    \item Gaussian kernel $K(\textbf{u})=e^{-\lVert \textbf{u} \rVert^2}$,
    \item Epanechnikov kernel $K(\textbf{u})=(1-\lVert \textbf{u} \rVert^2) \, \mathds{1}_{\{\lVert \textbf{u} \rVert\leq 1\}}$,
    \item triangular kernel $K(\textbf{u})=(1-\lVert \textbf{u} \rVert) \, \mathds{1}_{\{\lVert \textbf{u} \rVert\leq 1\}}$,
    \item quartic kernel $K(\textbf{u})=(1-\lVert \textbf{u} \rVert^2)^2  \, \mathds{1}_{\{\lVert \textbf{u} \rVert\leq 1\}}$,
    \item triweight kernel $K(\textbf{u})=(1-\lVert \textbf{u} \rVert^2)^3 \, \mathds{1}_{\{\lVert \textbf{u} \rVert\leq 1\}}$,
    \item tricube kernel $K(\textbf{u})=(1-\lVert \textbf{u} \rVert^3)^3 \, \mathds{1}_{\{\lVert \textbf{u} \rVert\leq 1\}}$.
\end{itemize}
Note that only the Gaussian kernel assigns a positive weight to the furthest neighbor (or neighbors) $(n_t)$ of $(l)$. A straightforward weighted $k$-NN estimator is to consider the geographical locations of the wind turbines and to base the Nadaraya-Watson estimators on the geographical distances between the wind turbines, i.e., $\lVert \textbf{z}^{(j)} - \textbf{z}^{(l)} \rVert=\sqrt{\left(\text{latitude}^{(j)}-\text{latitude}^{(l)}\right)^2+\left(\text{longitude}^{(j)}-\text{longitude}^{(l)}\right)^2}$. We refer to this benchmark as the ``location'' estimator.

\subsection{Wind farms as unweighted graphs}\label{sec:unweightedgraphs}
\subsubsection{Graphs}\label{sec:graph}
A graph $G$ is defined by a set of nodes \textcolor{blue}{(}or vertices\textcolor{blue}{)} $V=v_1,\dots,v_N$ and a set of edges $E$ between nodes. It is said to be undirected if there is no direction implied by an edge. Often when a vertex $v_i$ represents a data point $x_i$, two vertices $v_i$ and $v_j$ are connected if $x_i$ and $x_j$ are close. Let the wind farm be a graph $G=(V,E)$ where the set of nodes $V$ are the wind turbines, $\vert V \vert = N$, and the set of edges $E$ connecting two nodes are to be decided upon. We build our graph out of the layout of the wind farm, without considering any data points $x_i$. We start with an unweighted graph, i.e., the edges of the graph are unweighted. Let $\textbf{A}=(a_{ij})_{i,j=1,\dots,N}$ be the \textit{adjacency} matrix of the graph $G$. Unweighted edges (or simple-minded weights) means that $a_{ij}=1$ if vertices $v_i$ and $v_j$ are connected by an edge, $a_{ij}=0$ otherwise. Hence, all edges are assumed to have the same strength. Note that the diagonal of \textbf{A} is equal to zero, i.e., $a_{ii}=0 \ \forall \ i=1,\dots,N$, as we do not consider self-connections. Through the adjacency matrix \textbf{A}, each wind turbine is represented by the vector of size $N$ of its connections to the other wind turbines in the wind farm. Out of this representation, we wish to learn a low-dimensional embedding for each wind turbine that preserves the structure of the wind farm. 

\subsubsection{Laplacian eigenmaps}\label{sec:eigenmaps}
We are interested in spectral-graph embeddings, and in particular in Laplacian eigenmaps, which optimally preserve local neighborhood information and produce coordinate maps that are smooth functions over the original graph \cite{Belkin2003}. By trying to preserve local information in the embedding, the algorithm implicitly emphasizes the natural clusters in the data and closely relates to spectral clustering \cite{vonLuxburg2007}. We hope for the Laplacian eigenmaps to provide a smooth clustering of the wind turbines over the wind farm. Let \textbf{D} be the diagonal matrix associated with the graph $G$ whose entries are the degree of each node, i.e., $d_{ii}=\sum_j a_{ji}$. The matrix $\textbf{L}=\textbf{D}-\textbf{A}$ is called the Laplacian matrix of the graph and one gets eigenmaps by computing eigenvalues and eigenvectors for the generalized eigenvalue problem 
\begin{equation}
\label{eq:eigenpb}
    \textbf{L}\textbf{f} = \lambda \textbf{D}\textbf{f}.
\end{equation}
Let $\textbf{f}_0,\dots,\textbf{f}_{N-1}$ be the solutions of Equation \eqref{eq:eigenpb}, ordered according to their eigenvalues\textcolor{blue}{:} 
\begin{subequations}
    \begin{align}
    \textbf{L}\textbf{f}_0 &= \lambda_0  \textbf{D}\textbf{f}_0\textcolor{blue}{,}\\
    \textbf{L}\textbf{f}_1 &= \lambda_1  \textbf{D}\textbf{f}_1\textcolor{blue}{,}\\
    &... \\
    \textbf{L}\textbf{f}_{N-1} &= \lambda_{N-1}  \textbf{D}\textbf{f}_{N-1}\textcolor{blue}{,}\\
    0=\lambda_0 \leq &\lambda_1 \leq \dots \leq \lambda_{N-1}.
\end{align}
\end{subequations}
We leave out the constant eigenvector $\textbf{f}_0$ corresponding to eigenvalue 0 and use the next $r$ eigenvectors for embedding each wind turbine $v_i$ in a $r$-dimensional Euclidean space:
\begin{equation}
    \textbf{z}^{i} = (\textbf{f}_1(v_i),\dots,\textbf{f}_r(v_i)).
\end{equation}
The embedding $\textbf{z}^i$ is a new representation of the wind turbine $v_i$. Distances $\lVert \textbf{z}^{i} - \textbf{z}^{j} \rVert$ for every pair of wind turbines $(v_i,v_j)$ can now be computed once and for all\textcolor{blue}{,} but the number of components we keep in $\textbf{z}^i$ needs to be decided upon. For each missing data point\textcolor{blue}{,} the kernel function will then adjust the weights at each time $t$ depending on the set of $n_t$ data points that are available from other wind turbines.

\subsection{Wind farms as weighted graphs}\label{sec:weightedgraphs}
\subsubsection{Weighting the original graph}
The representations we get out of the unweighted graph embed the structure of the wind farm only and can be used without having any other data but the map of the wind farm. The unweighted graph can be seen as a stationary a priori component of the relationship between the wind turbines, which comes from the location of the wind turbines inside the wind farm. It may be completed with an online component coming from the time series we are interested in. We propose to keep the structure of the unweighted graph $G$ and to move from unweighted to weighted edges. Say we are interested in $X_t^i$, the power generation of wind turbine $v_i$ at time $t$ normalized by the nominal capacity of the wind turbine. We have $X_t^i \in [0,1]$ for $t=1,\dots,T$, $i=1,\dots,N$. An edge between two wind turbines is to be weighted according to how similar these wind turbines are. Working with power generation, this translates to the similarity between their productions at time $t$. Let $x_t^i$, resp. $x_t^j$, be the observed power generation of wind turbine $v_i$, resp. $v_j$, at time $t$. A simple and intuitive choice for the similarity between $x_t^i$ and $x_t^j$ is $s_t(i,j)=1-\lvert x_t^i - x_t^j \rvert$, $s_t(i,j) \in [0,1]$. Note that if two wind turbines that are not connected in graph $G$ happen to have very similar power generation values at time $t$, they remain unconnected. By doing so, we enforce a stationary a priori on the relationship between the wind turbines out of the structure of the wind farm, but not only: we also keep a sparse adjacency matrix, which now depends on $t$. Indeed, we have $\textbf{A}_t=(a_{t,ij})_{i,j=1,\dots,N}$, where $a_{t,ij}=s_t(i,j)$ if vertices $v_i$ and $v_j$ are connected by an edge, $a_{t,ij}=0$ otherwise. This implies to solve the generalized eigenproblem \eqref{eq:eigenpb} at each time step $t$, which can be done rather easily when dealing with sparse matrices $\textbf{A}_t$. 

\subsubsection{Online, changing graphs} \label{sec:onlinegraph}
Because of the way $G$'s edges are now weighted, it can happen that two wind turbines that are a priori connected get disconnected because $s_t(i,j)=0$ for some time $t$. In such a case, we can get a graph that is not \textit{connected} anymore, i.e., we cannot travel through the whole graph from any point in the graph. To compute eigenmaps the generalized eigenproblem \eqref{eq:eigenpb} must be solved for a connected graph, which ensures $\text{rank}(\textbf{D})=N$ and there are $N$ eigenvalues, or $\lambda(\textbf{L},\textbf{D})$ may be finite, empty or infinite \cite{Golub2013}. Therefore, when a graph has several components, the algorithm for computing eigenmaps consists of solving the generalized eigenproblem \eqref{eq:eigenpb} for each connected component, which we will do for components with at least three wind turbines. When a wind turbine splits from the graph on its own, it is straightforward to derive its production value from its neighbors as typically a wind turbine $v_i$ gets disconnected at time $t$ because $x_t^i=0$ and $x_t^j=1$ for all its neighbors $v_j$. When two wind turbines get disconnected together, the similarity between the two is usually high enough to replace the record that is missing with the one that is available. Since we are online and in high dimension, we need to be able to automatically detect when the graph is not connected anymore, and what are its connected components. Let $\textbf{L}_{rw}=\textbf{D}^{-1}\textbf{L}=\textbf{I}-\textbf{D}^{-1}\textbf{A}$ be the so-called normalized graph Laplacian, which is the graph Laplacian we use to compute eigenmaps. We recall a basic yet very useful property of this graph Laplacian \cite{vonLuxburg2007}, that makes it easy to derive the connected components of $G$ at time $t$ if the multiplicity of the eigenvalue 0 of $\textbf{L}_{rw,t}$ becomes higher than 1. Let $G$ be an undirected graph with non-negative weights. The multiplicity $d$ of the eigenvalue 0 of the graph Laplacian $\textbf{L}_{rw}$ equals the number of connected components $A_1,\dots,A_d$ in the graph and the eigenspace of 0 is spanned by the indicator vectors $\mathds{1}_{A_i}$ of those components.

So far we have assumed the similarities $s_t(i,j)$ to be known. Because our application is the imputation of missing values, the true distances are actually known at time $t$ among the wind turbines for which data points are available, but they are not for the edges involving  wind turbines for which the record is missing and we need to replace them with estimates. To avoid having to specify a model for the similarities between all individual time series, we place ourselves in the online learning framework. The goal in this learning paradigm is to guess a sequence of numbers as precisely as possible, when the data are chosen by an adversary rather than generated stochastically \cite{CesaBianchi2006,Orabona2022}. In our framework this translates as the following repeated game: in each round $t=1,\dots,T$, for each similarity between two connected wind turbines $s_t(i,j)$, an adversary chooses a real number in $[0,1]$ and keeps it secret; we try to guess the real number, choosing $\Hat{s}_t(i,j)$; the adversary number is revealed and we pay the squared difference $(s_t(i,j) - \Hat{s}_t(i,j))^2$. Online learning is appealing from both a theoretical and practical point of view because a lot of problems can be described as such a repeated game, which does not require strong assumptions to offer nice theoretical guarantees. One shall note that the last step of the repeated game when the adversary number is revealed does not happen if some of the data are missing. Therefore, in order to account for the possibility of missing data, the game is slightly modified and we pay $(s_t(i,j) - \Hat{s}_t(i,j))^2$ only if $s_t(i,j)$ is revealed \cite{Anava2015}. Using the notation $\mathds{1}_{\{s_t(i,j)\}}$ as the indicator of the event $\{s_t(i,j) \ \text{is revealed}\}$, we pay now $(s_t(i,j) - \Hat{s}_t(i,j))^2\mathds{1}_{\{s_t(i,j)\}}$. In such a missing data, convex framework, an online strategy with good theoretical guarantees is the lazy version of OGD \cite{Hazan2021}, which is applied to our problem in Algorithm \ref{alg1}, where $\Pi_{[0,1]}$ is the projection back to $[0,1]$. We also consider the best constant strategy, i.e, the strategy that minimizes $\sum_{t=1}^T (s_t(i,j) - \Hat{s}_t(i,j))^2\mathds{1}_{\{s_t(i,j)\}}$, which is just choosing $\Hat{s}_t(i,j)$ to be the average similarity $\sum_{t=1}^T s_t(i,j)\mathds{1}_{\{s_t(i,j)\}} \big/ \sum_{t=1}^T \mathds{1}_{\{s_t(i,j)\}}$ in each round $t$.
\begin{algorithm}[!h]
\caption{}\label{alg1}
\begin{algorithmic}
\Require learning rate $\eta$
\State Set $\Hat{s}_1(i,j)=y_1=1$.
\For{$t=1,\dots,T$}
\State Play $\Hat{s}_t(i,j)$ and occur cost $\displaystyle f_t(\Hat{s}_t)= (s_t(i,j) - \Hat{s}_t(i,j))^2\mathds{1}_{\{s_t(i,j)\}}$.
\State Update and project:
\begin{align*}
    y_{t+1} &= y_{t} - \eta \nabla f_t(\Hat{s}_t)\\
    \Hat{s}_{t+1}(i,j) &= \Pi_{[0,1]}(y_{t+1})
\end{align*}
\EndFor
\end{algorithmic}
\end{algorithm}

By choosing the online learning framework, we do not only free ourselves from any model assumption, we also allow our weighted-graph estimator to be of use as soon as there are data points for some of the wind turbines. At the beginning of a wind farm's life, when not all wind turbines are on yet, the estimator can work with a graph $G$ being restricted to the operational wind turbines. As soon as a wind turbine gets started, Algorithm \ref{alg1} can be run to estimate the similarities that might be missing between this wind turbine and its neighbors. The theory underlying the lazy version of OGD ensures that we minimize our \textit{regret} $\mathcal{R}_t(i,j)$, i.e., the quantity that measures how much our algorithm regrets for not sticking to the optimal choice in hindsight after $t$ iterations \cite{Orabona2022, Hazan2021}:
\begin{equation}
    \mathcal{R}_t(i,j) = \sum_{k=1}^t\left(\Hat{s}_k(i,j)-s_k(i,j)\right)^2 - \min_{s \in [0,1]}\sum_{k=1}^t (s-s_k(i,j))^2.
\end{equation} 

\subsection{Computational complexity of the estimators}
Let $T_{mv}$ be the number of time steps $t=1,\dots,T$ such that $n_t < N$. The computational complexity of the estimators we have introduced is available in Table \ref{tab2}.
\begin{table*}[!h]%
\centering %
\caption{Estimators for the $n_t$-NN imputation of power generation's missing values, along with their computational complexity.}\label{tab2}%
\begin{tabular*}{\textwidth}{@{\extracolsep\fill}llll@{\extracolsep\fill}}
\toprule
\textbf{Estimator} & $k$\textbf{-NN} & \textbf{Graph} & \textbf{Computational complexity}\\
\midrule
Naive & Unweighted & No & $O(T_{mv}N)$\\
Location & Weighted & No & $O(N^2+T_{mv}N)$\\
Unweighted-graph & Weighted & Yes, unweighted & $O(N^3+T_{mv}N)$\\
Weighted-graph & Weighted & Yes, weighted & $O(T_{mv}N^3)$\\
\bottomrule
\end{tabular*}
\end{table*}
It increases alongside the complexity of the method at hand. The naive estimator only requires to average over the available records at each time step $t \in 1,\dots,T_{mv}$. Before computing a weighted average, the location estimator requires to first compute, once and for all, the geographical distances between wind turbines. Similarly, the unweighted-graph estimator requires to first solve the generalized eigenvalue problem in Equation \eqref{eq:eigenpb}. The weighted-graph estimator is the most demanding in terms of time complexity since it requires to solve the generalized eigenvalue problem at each time step $t \in 1,\dots,T_{mv}$ before computing a weighted average. On the other hand, it is the only one that accounts for information provided by collected power generation data.

\section{Application and case-study: Westermost Rough}\label{sec:results}
\begin{figure*}[!h]
\centerline{\includegraphics[width=0.6\textwidth]{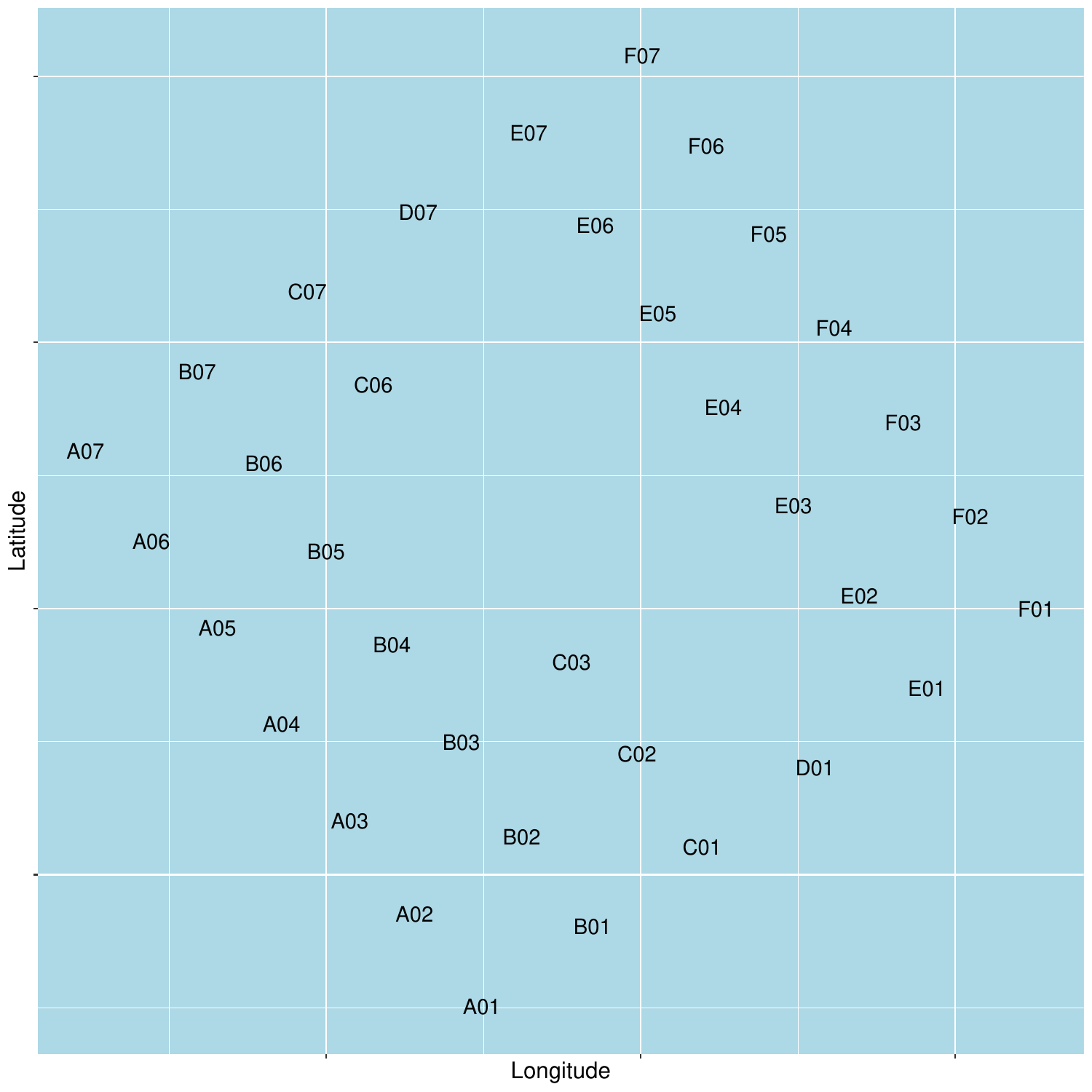}}
\caption{Position and name of the 35 wind turbines in Westermost Rough offshore wind farm (UK).\label{fig1}}
\end{figure*}

We apply the method presented in Section \ref{sec:method} to a real use case, the Westermost Rough offshore wind farm. Westermost Rough is located near the Eastern coast of the United Kingdom and totals 35 wind turbines that are placed according to a grid pattern. A representation of the wind farm through the position and name of its wind turbines is available in Figure \ref{fig1}.
The pattern is rather usual and the number of wind turbines high enough to support our method, but not too high for us to deliver a detailed and visual analysis. Through this example, we wish to deliver good practices and to emphasize challenges that generalize to bigger and/or more complicated wind farms. Along with the exact position of the wind turbines, we have data records over two years, from January 1, 2016 to December 31, 2017, at a temporal resolution of every 10 minutes. We will focus on the graph representations in Section \ref{sec:WMR.graph} and will apply the method to power generation imputation in Section \ref{sec:power}.

\subsection{The Westermost Rough graph representations} \label{sec:WMR.graph}
\begin{figure*}[t]
\centerline{\includegraphics[width=0.6\textwidth]{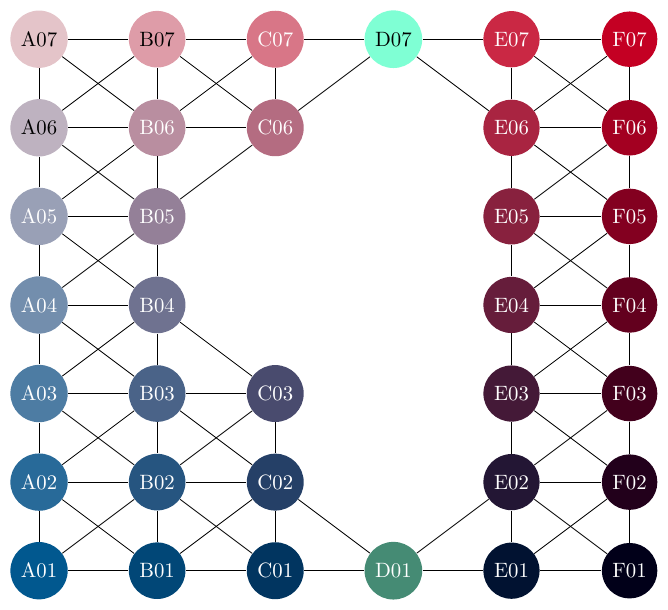}}
\caption{Graph of the Westermost Rough offshore wind farm with color code.\label{fig2}}
\end{figure*}

When constructing a graph, one's objective is to model the local neighborhood relationships and we choose to connect wind turbines that are actual neighbors, i.e., there is no other wind turbine nor empty space between them. The corresponding graph is presented in Figure \ref{fig2}. 
This is an important step for which there is no absolutely right choice and one should keep in mind that choosing an appropriate a priori graph matters to the results. For example, in the case of Westermost Rough, one could choose not to connect wind turbines that are neighbors through a diagonal. How did we choose this graph? When looking at the exact locations, it turns out that the grid pattern is not exact and in reality two wind turbines on what looks like a diagonal might be closer than two wind turbines on an horizontal line. Also, the direction of the wind might matter more than the real distances and, last but not least, the more neighbors for a wind turbine may be the better.

\begin{figure*}[b]
\centerline{\includegraphics[width=\textwidth]{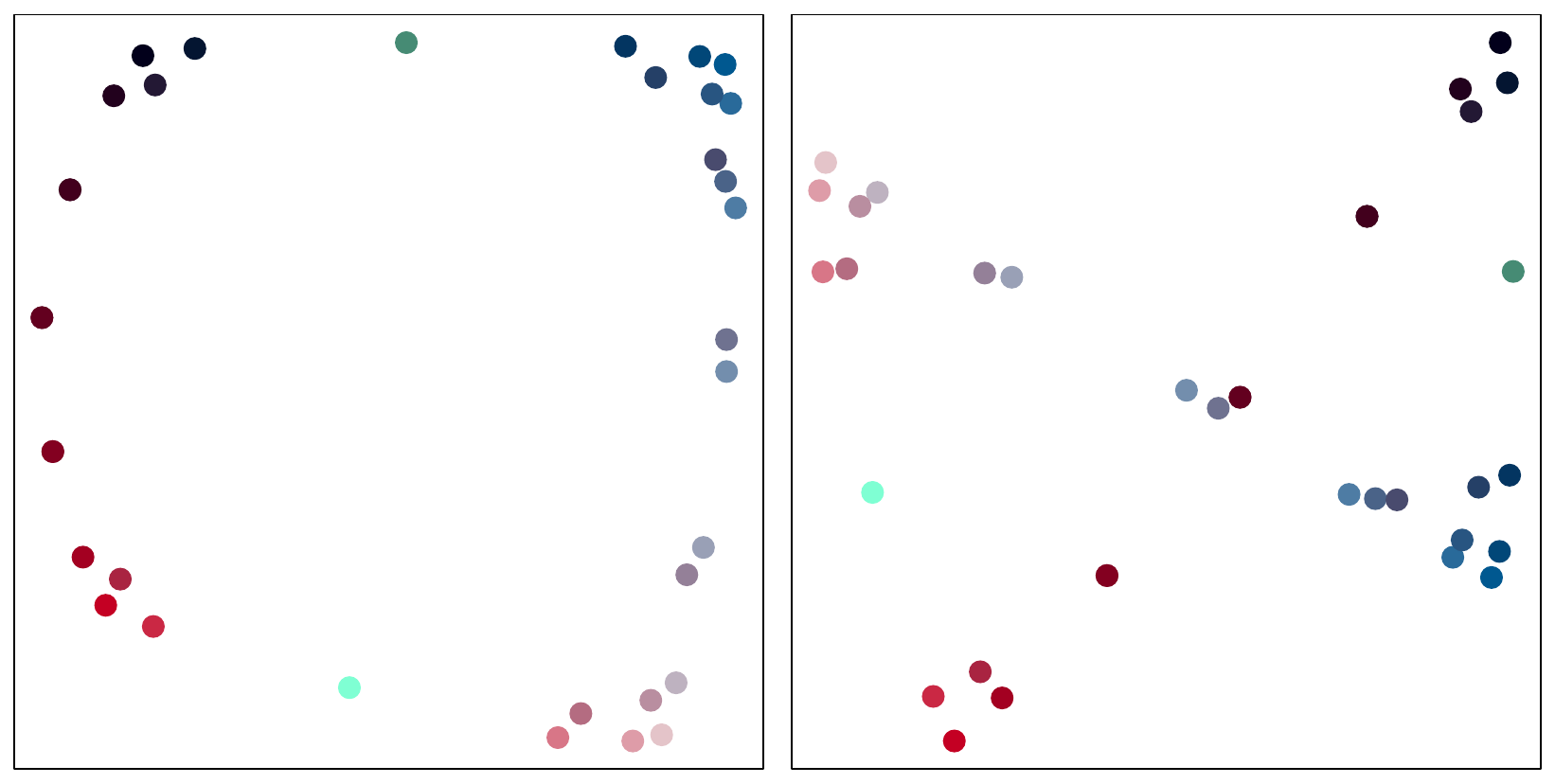}}
\caption{Representation of the wind turbines according to dimensions 1 and 2 (left) and dimensions 2 and 3 (right).\label{fig3}}
\end{figure*}

Now that we have decided upon our reference graph $G=(V,E)$, we start with an unweighted version of $G$, i.e., the edges $E$ have 0/1 weights only. We compute the corresponding Laplacian matrix $\textbf{L}=\textbf{D}-\textbf{A}$ and solve the generalized eigenvalue problem in \eqref{eq:eigenpb}. We leave out the (first) constant eigenvector and get an embedding $\textbf{z}^i$ for every wind turbine $v_i$. These embeddings are of maximal size 34, i.e., $N$ minus the first constant eigenvector $\textbf{f}_0$. In Figure \ref{fig3}\textcolor{blue}{,} we show the corresponding representations of the wind turbines according to eigenvectors $\textbf{f}_1$ and $\textbf{f}_2$ (left) and eigenvectors  $\textbf{f}_2$ and $\textbf{f}_3$ (right).
On these eigenmaps, the wind turbines are smoothly clustered together over the wind farm according to their position, where both the local neighborhoods and the whole structure of the wind farm are accounted for. These plots provide nice insight about the representations learned from running the eigenmap algorithm over the unweighted graph, and how they embed the local and global geometry of the wind farm's layout in a lower dimension.

We focus on imputing missing data for wind power generation. Hence, we can weight the edges $E$ of $G$ using the methodology and the similarity described in Section \ref{sec:weightedgraphs}, where $s_t(i,j)=1-\lvert x_t^i - x_t ^j \rvert$ is the similarity at time $t$ between the power generations $x_t^i$ and $x_t^j$ of two connected wind turbines $v_i$ and $v_j$. Working with weighted graphs whose weights vary over time, the eigenmap algorithm needs to be run, and we get different embeddings, at every time $t$. As mentioned in Section \ref{sec:onlinegraph}, the graph itself can change, if an edge's weight becomes 0 at time $t$. As for the representations learned by the eigenmap algorithm, the clusters are still dependent upon the original graph $G$, i.e., upon the geography of the wind farm, but the distances between the embeddings may now change depending on an edge's weight $s_t(i,j)$.

\subsection{Imputation of power generation values}\label{sec:power}
\subsubsection{Evaluation setup}
In the methodology we propose, we do not estimate any parameters and try to minimize the number of decisions that need to be made. Nonetheless, when moving to imputation, we need to decide on a few user-specified parameters (or hyperparameters), which are:
\begin{itemize}
    \item the number of dimensions $r$ we keep for the embeddings $\textbf{z}^i$ we get out of the unweighted graph or the embeddings $\textbf{z}^i_t$ we get out of the weighted graphs,
    \item the kernel function that turns the distances between these embeddings into weights for the weighted $k$-NN imputation,
    \item the learning rate $\eta$ in Algorithm \ref{alg1} if we use weighted graphs and lazy OGD.
\end{itemize}
Recall that we have two years of data records over the Westermost Rough wind farm, 2016 and 2017. We split this dataset into two sets: a validation set, the first year of data, 2016, for deciding upon the hyperparameters; a test set, the second year, 2017, for evaluating our method compared to the naive Nadaraya-Watson estimator that is our current reference and to the location Nadaraya-Watson estimator that is a standard weighted $k$-NN estimator. We evaluate the different estimators through the RMSE
\begin{equation}
\label{eq:RMSE}
    \sqrt{\frac{1}{T_i} \sum_{t=1}^{T_i} \left(x_t^{i} - \sum_{j=1}^{n_t} w_t^{(ji)}x_t^{(j)}\right)^2},
\end{equation}
which we will compute for each wind turbine $v_i$, where $T_i$ is the number of records for which $x_t^i$ is available. We compute the RMSE in Equation \eqref{eq:RMSE} for two different setups: 
\begin{itemize}
    \item a setup we call ``incomplete'', where we compute \eqref{eq:RMSE} for all data records that include $x_t^{i}$, no matter what other data records are available at time $t$;
    \item a setup we call ``complete'', where we compute \eqref{eq:RMSE} on complete data records only, i.e., such that we have $x_t$ for all $N$ wind turbines.
\end{itemize}
The validation set consists of 52,669 records, 23,499 records being complete records. The test set consists of 52,549 records, 29,560 records being complete records. The incomplete setup enables us to evaluate the quality of the estimates for each wind turbine, taking into account the reality of the availability of other records, while the complete setup measures an ideal quality of the estimate, in the sense that we assume all the other records to be available. Note that to simulate our own missing values is not really an option on this dataset, as we only have nearly half of the records that are complete records, and they are unlikely to be successive records, breaking down the dynamics of the time series. 

\subsubsection{Hyperparameter selection}
The estimators we are evaluating are
\begin{itemize}
    \item the naive estimator: Nadaraya-Watson estimator with a naive kernel, which comes down to equal weights for all available data records;
    \item the location estimator: Nadaraya-Watson estimator that assigns weights depending on the geographical distances between the wind turbines;
    \item the unweighted-graph estimator: Nadaraya-Watson estimator that assigns weights depending on the distances between embeddings obtained from an unweighted graph $G$;
    \item the weighted-graph estimator: Nadaraya-Watson estimator that assigns weights depending on the distances between embeddings obtained from online weighted graphs $G_t$.
\end{itemize}

On the validation set, the location estimator gives the best results when using a triweight kernel. The improvements achieved on the RMSE compared to using a naive kernel are listed in Table \ref{tab3}, as averages over the wind farm, along with the standard deviation depending on the wind turbine.
\begin{table*}[!h]%
\centering %
\caption{Average improvement by kernel on the validation set for the location estimator.\label{tab3}}%
\begin{tabular*}{\textwidth}{@{\extracolsep\fill}lcccccc@{\extracolsep\fill}}
\toprule
\textbf{Setup} & \textbf{Gaussian} & \textbf{Epanech} & \textbf{Triangular} & \textbf{Quartic} & \textbf{Triweight} & \textbf{Tricube}\\
\midrule
Incomplete & 1.81\% (0.95) & 2.58\% (1.42) & 3.45\% (1.96) & 3.83\% (2.27) & 4.52\% (2.87) & 3.66\% (2.20) \\
Complete & 2.10\% (1.08) & 3.00\% (1.58) & 4.07\% (2.20) & 4.55\% (2.51) & 5.44\% (3.13) & 4.34\% (2.41) \\
\bottomrule
\end{tabular*}
\end{table*}
The unweighted-graph estimator is always better using a dimension $r=2$ for the embeddings. While in the complete setup to use a triweight kernel gives the best results, there is less difference between using a triweight and a quartic kernel in the incomplete setup. The improvements achieved on the RMSE depending on the dimension $r$ and the kernel are listed in Table \ref{tab4}. The improvements depending on the dimension are averaged over the wind turbines and the different kernels, while the improvement depending on the kernel are shown for $r=2$. 
\begin{table*}[!h]%
\centering %
\caption{Average improvement by dimension and by kernel on the validation set for the unweighted-graph estimator.\label{tab4}}%
\begin{tabular*}{\textwidth}{@{\extracolsep\fill}lcccccc@{\extracolsep\fill}}
\toprule
\textbf{Setup} & $r=1$ & $r=2$ & $r=3$ & $r=4$ & $r=5$\\
\midrule
Incomplete & 2.40\% (3.31) & 4.21\% (3.18) & 3.60\% (2.92) & 3.61\% (3.08) & 3.03\% (2.82)\\
Complete & 2.73\% (3.95) & 5.17\% (3.52) & 4.55\% (3.27) & 4.51\% (3.36) & 3.83\% (3.17)\\
\bottomrule
\toprule
\textbf{Setup} & \textbf{Gaussian} & \textbf{Epanech} & \textbf{Triangular} & \textbf{Quartic} & \textbf{Triweight} & \textbf{Tricube}\\
\midrule
Incomplete & 2.53\% (1.30) & 3.99\% (2.42) & 4.52\% (2.96) & 4.75\% (3.51) & 4.79\% (4.10) & 4.70\% (3.55)\\
Complete & 3.04\% (1.51) & 4.85\% (2.67) & 5.52\% (3.25) & 5.84\% (3.85) & 5.95\% (4.45) & 5.81\% (3.89)\\
\bottomrule
\end{tabular*}
\end{table*}

For the weighted-graph estimator, we need to use estimated similarities on the edges between the wind turbine we are testing and its neighbors in the complete setup, and on all edges involving missing data records in the incomplete setup. We can choose what estimated similarity to use without running the eigenmap algorithm by looking at the loss we pay on the validation set:
\begin{equation}
\label{eq:loss}
    \sum_{t=1}^{T_\text{val}} l_t(i,j) = \sum_{t=1}^{T_\text{val}} (s_t(i,j) - \Hat{s}_t(i,j))^2\mathds{1}_{\{s_t(i,j)\}}.
\end{equation}
In the online learning paradigm, the loss in \eqref{eq:loss} is a reference when computed for the best constant strategy 
\begin{equation}
    \Hat{s}(i,j)=\sum_{t=1}^{T_\text{val}} s_t(i,j)\mathds{1}_{\{s_t(i,j)\}} \bigg/ \sum_{t=1}^{T_\text{val}} \mathds{1}_{\{s_t(i,j)\}}.
\end{equation}
The difference between the loss of any other strategy and the loss of the best constant strategy is known as the regret (of not sticking to the best choice in hindsight, as introduced in Section \ref{sec:method}) and is what theoretical results for online learning mostly focus on, i.e., ensuring the regret is (nicely) bounded when using a specific strategy. When computing the regret of lazy OGD on the validation set for different values of the learning rate $\eta$, it is clear that lazy OGD provides far better results than the best constant strategy. This is a confirmation that the similarities between wind power time series are nonstationary. An optimal learning rate can be computed from theoretical results on lazy OGD. Let $D$ be the diameter of the support $\mathcal{K}$ of the loss function $l_t$ we pay in each round $t$, $B$ an upper bound to the norm of the gradients of $l_t$ and $T$ the number of rounds. The regret of lazy OGD is best bounded by taking $\eta=D (B\sqrt{T})^{-1}$ \cite{Hazan2021}. We have $\mathcal{K}=[0,1]$, $\displaystyle D=\sqrt{\max_{s_t,\Hat{s}_t \in \mathcal{K}}(s_t(i,j)-\Hat{s}_t(i,j))}=1$ and $B=2$ as 
\begin{equation}
    \lVert \nabla l_t(i,j) \rVert=\lVert -2(s_t(i,j)-\Hat{s}_t(i,j))\mathds{1}_{\{s_t(i,j)\}}\rVert \leq 2. 
\end{equation}
Since we deal with missing data, we get $\eta=\left( 2\sqrt{\sum_{t=1}^{T_\text{val}}\mathds{1}_{\{s_t(i,j)\}}} \right)^{-1}$ \cite{Anava2015}. The learning rate is usually set to decrease over time when one is interested in converging to an optimal solution. Since we want to track a time-varying quantity, a standard approach is to rather choose the learning rate to be constant. Therefore, we take $\eta=0.5$. This is empirically verified on our validation set as better regrets are obtained for constant learning rates in $[0.3,0.5]$. In the case where $\eta=0.5$, lazy OGD comes down to what is known as persistence in time series forecasting without missing values, which is simply to use the last observed value. When using lazy OGD with $\eta=0.5$, the best kernel is again the triweight kernel in the complete setup, but the best results are now obtained using a dimension $r=4$ for the embeddings. In the incomplete setup, the best estimator is also the one that uses a triweight kernel and a dimension $r=4$ for the embeddings. The improvements achieved on the RMSE depending on the dimension $r$ and the kernel are listed in Table \ref{tab5}. The improvements depending on the dimension are averaged over the wind turbines and the different kernels, while the improvement depending on the kernel are shown for $r=4$.
\begin{table*}[!h]%
\centering %
\caption{Average improvement by dimension and by kernel on the validation set for the weighted-graph estimator.\label{tab5}}%
\begin{tabular*}{\textwidth}{@{\extracolsep\fill}lcccccc@{\extracolsep\fill}}
\toprule
\textbf{Setup} & $r=1$ & $r=2$ & $r=3$ & $r=4$ & $r=5$\\
\midrule
Incomplete & 4.95\% (3.47) & 7.09\% (3.21) & 6.80\% (3.23) & 7.40\% (3.79) & 6.85\% (3.44)\\
Complete & 5.63\% (4.60) & 8.15\% (4.34) & 7.66\% (4.21) & 8.13\% (4.41) & 7.45\% (4.15)\\
\bottomrule
\toprule
\textbf{Setup} & \textbf{Gaussian} & \textbf{Epanech} & \textbf{Triangular} & \textbf{Quartic} & \textbf{Triweight} & \textbf{Tricube}\\
\midrule
Incomplete & 3.31\% (1.00) & 5.72\% (2.08) & 7.27\% (2.61) & 8.94\% (3.45) & 10.00\% (4.26) & \ 9.19\% (3.67)\\
Complete & 3.61\% (1.39) & 6.31\% (2.71) & 7.96\% (3.25) & 9.82\% (4.06) & 10.93\% (4.98) & 10.15\% (4.34)\\
\bottomrule
\end{tabular*}
\end{table*}

\subsubsection{Results on the test set}
\begin{figure*}[!h]
\centerline{\includegraphics[width=\textwidth]{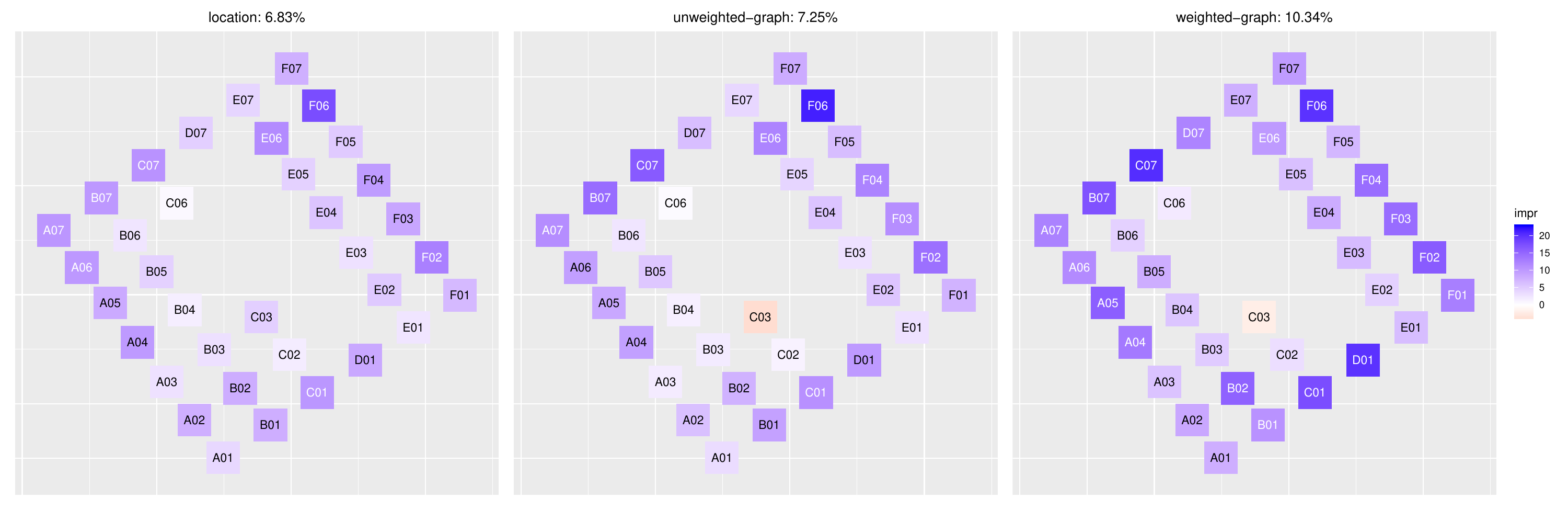}}
\caption{Improvements over the naive estimator by wind turbine in the complete setup.\label{fig4}}
\end{figure*}

\begin{figure*}[!h]
\centerline{\includegraphics[width=\textwidth]{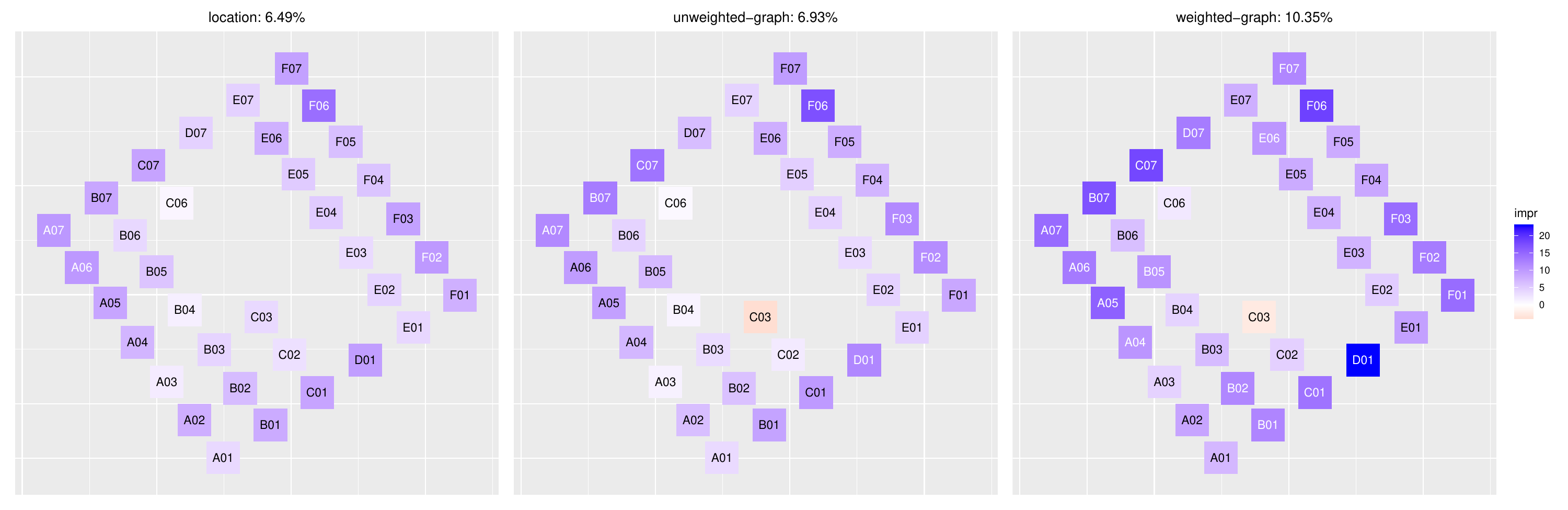}}
\caption{Improvements over the naive estimator by wind turbine in the incomplete setup.\label{fig5}}
\end{figure*}

Out of the results on the validation set, we compute the estimators over the test set using a triweight kernel. For the unweighted-graph estimator, we use embeddings of dimension 2. For the weighted-graph estimator, we use embeddings of dimension 4 and lazy OGD with $\eta=0.5$. The results on the test set in the complete, resp. incomplete, setup are given in Table \ref{tab6}, resp. in Table \ref{tab7}. They are averaged over the wind turbines. 
\begin{table*}[!h]%
\centering %
\caption{Results on the test set in the complete setup.\label{tab6}}%
\begin{tabular*}{\textwidth}{@{\extracolsep\fill}lcccc@{\extracolsep\fill}}
\toprule
\textbf{Estimator} & \textbf{avg RMSE (sd)}  & \textbf{ avg impr./naive}  & \textbf{best impr./naive ($v_i$)}  & \textbf{worst impr./naive ($v_i$)} \\
\midrule
naive & 12.32\% (2.86)  & -  & -  & -   \\
location & 11.55\% (3.02)  & \ 6.83\%  & 17.69\% (F06)  & \ 0.65\% (C06)   \\
unweighted-graph & 11.50\% (3.06)  & \ 7.25\%  & 21.36\% (F06)  & -4.10\% (C03) \\
weighted-graph & 11.11\% (2.95)  & 10.34\%  & 20.50\% (C07)  & -2.11\% (C03)  \\
\bottomrule
\end{tabular*}
\end{table*}
\begin{table*}[!h]%
\centering %
\caption{Results on the test set in the incomplete setup.\label{tab7}}%
\begin{tabular*}{\textwidth}{@{\extracolsep\fill}lcccc@{\extracolsep\fill}}
\toprule
\textbf{Estimator} & \textbf{avg RMSE (sd)}  & \textbf{ avg impr./naive}  & \textbf{best impr./naive ($v_i$)}  & \textbf{worst impr./naive ($v_i$)} \\
\midrule
naive & 12.14\% (2.35)  & -  & -  & -   \\
location & 11.39\% (2.44)  & \ 6.49\%  & 14.45\% (F06)  & \ 0.94\% (C06)   \\
unweighted-graph & 11.34\% (2.47)  & \ 6.93\%  & 17.31\% (F06)  & -4.01\% (C03) \\
weighted-graph & 10.92\% (2.42)  & 10.35\%  & 23.20\% (D01)  & -2.59\% (C03)  \\
\bottomrule
\end{tabular*}
\end{table*}
The improvement over the naive estimator is plotted by wind turbine in Figure \ref{fig4}, resp. in Figure \ref{fig5}, for each estimator. 
Accounting for the geographical distance between a wind turbine and its neighbors already improves the estimates for every wind turbine compared to a naive estimator that assigns the same weight to all neighbors. With additional information about the distances between the neighbors themselves, through the structure of the wind farm, the estimators perform better on average over the wind farm, but there is more variability depending on the wind turbine. In particular, the power generation of wind turbine C03 is better approximated by a naive estimator. 

The improvement we get from using the weighted-graph estimator is significant overall, for both complete and incomplete setups, and more stable over wind turbines. The weighted-graph estimator performs better imputation for wind turbines that are on the outer parts of the wind farm, up to more that 20\% for some of them. As a reference, if we use the real similarities for the edges in the complete setup instead of the estimates from lazy OGD, the overall improvement over the naive estimator is 11.21\%, compared to 10.34\% with the estimates. There is still room for improving the estimated similarities, for example by taking into account exogenous information such as the wind direction or speed, but the lazy OGD already performs pretty well.

Finally, note that on the test set 43.76\% of the records are not complete. Most of these incomplete records miss wind power generation data for one (69.57\%), two (23.22\%) or three wind turbines (6.79\%). The case when only one record is missing is well described by our complete setup, where we remove the record of only one wind turbine and look into how well the missing data point is imputed by the estimator, depending on which turbine it has been removed for. 

\subsection{Discussion}\label{sec:discussion}
First, let us acknowledge that when working at an aggregated level, over large wind farms, if in most cases only one record is missing at time $t$ out of $N$, averaging over the remaining $N-1$ records seems fair enough, at least for Westermost Rough. However, this might not be the case for wind farms with less conventional layouts. Accounting for the geographical distances when computing the weights of a weighted average can already lead to significant improvements. The additional information of the structure of the wind farm gave better estimators on average, but at the cost of more instability, by degrading the imputation for some wind turbines. In particular, this was the case for wind turbine C03, which does not benefit from employing more advanced estimators, although this degradation was limited by including information about the power generation through a weighted graph. Wind turbine C03 has a very central position, which supports the choice of an estimator with equal weights. It might also be that C03's closer neighbors show a rather different behaviour and do not help much in estimating its power generation, e.g., because of wave effects. An option may be to consider a different graph $G$, which would link C03 to another wake-affected turbine, such as C06. This is directly related to our comment in Section \ref{sec:WMR.graph} about the importance of deciding on a graph. Note that we have restrained ourselves to the choice of a common kernel over the wind farm. If choosing a different dimension $r$ for the embeddings would not make much sense, to choose a different kernel for different wind turbines seems rather appropriate depending on the position of the wind turbine in the wind farm. However, moving into that direction, we would rather opt for replacing the Nadaraya-Watson estimators with an algorithm that can compute optimal weights efficiently and adaptively for each data point we wish to estimate, out of the distances between the wind turbines \cite{Anava2016}. 

If one is interested in individual signals, the methodology we propose can make much more difference, since for some wind turbines the graph-based estimators improve the power generation estimates by more than 20\%. We have focused here on what is known as single imputation as we have tried to impute missing entries as accurately as possible, which gives us only one completed dataset. Multiple imputation on the other hand consists in predicting $M$ different values for each missing data point and provides $M$ imputed datasets. Multiple imputation is usually preferred, for inference tasks in particular, as it ensures that the variance is properly accounted for. We want to emphasize that $k$-NN nicely enable to move to a probabilistic framework, when one is interested in the distributions of the time series, since a weight $w^{(jl)}$ can be seen as the probability of the missing data point $x^{(l)}$ to take the value of the available data point $x^{(j)}$. Let $\delta_{x^{(j)}}(x)$ denote the Dirac delta mass located at $x^{(j)}$. Instead of using a weighted average as a point estimate of $x^{(l)}$, we can assume $x^{(l)}$ to be distributed according to the empirical measure 
\begin{equation}\label{eq:empirical}
    \Hat{\pi}(x)=\sum_{j=1}^{n} w^{(jl)} \delta_{x^{(j)}}(x),
\end{equation}
and simply sample from \eqref{eq:empirical}, i.e., select $x^{(j)}$ with probability $w^{(jl)}$.

The location and unweighted-graph estimators can be very useful early on in the life of a wind farm, as they do not require any data points. The weighted-graph estimator does use data points, but can start as soon as there are data points for some wind turbines, as it does not require any model assumption nor estimation. The a priori we base our method on, by building the graph upon the structure of the wind farm, can be seen both as a pro and a con of the method, since it enforces proximity between wind turbines depending on their position inside the wind farm. This can bring robustness if it is appropriate, or instability when it is not, as we have seen with wind turbine C03. We have already mentioned this could be mitigated by moving from Nadaraya-Watson estimators to optimal weights thanks to an efficient algorithm \cite{Anava2016}, or by choosing a different graph. When using data points, that is weighted graphs, an alternative would be to remove this assumption and start from what is known as a \textit{complete} graph, where all pairs of wind turbines are connected by an edge. Then, the complete graph would evolve online depending on the similarity on the edges as in section \ref{sec:onlinegraph}. However, this would require to monitor $N(N-1)/2$ similarities and to perform eigendecomposition on matrices that are unlikely to be sparse anymore.

\section{Conclusion and future directions}\label{sec:conclusion}
%"The best thing to do with missing values is not to have any" Gertrude Mary Cox ahahah
From the intuitive practice of averaging over a wind farm in order to deal with missing data points, we have focused on weighted $k$-NN imputation for wind power generation and dealt not only with distances between a wind turbine and its neighbors but also with distances between the neighbors themselves by learning graph representations. Weighting the graph edges with the similarity at time $t$ between two data points from two time series and using spectral graph theory has enabled us to compute online representations that could adapt to changes in the relationship between a wind turbine and its neighbors, typically when a wind turbine is not producing. The methods we have introduced may be applied to perform imputation for other quantities of interest over the wind farm. However, not all quantities would share the nice feature of normalized power generation, namely to belong to the unit interval $[0,1]$. If not, one would need to rethink Algorithm \ref{alg1}, which estimates the missing similarities when working with an online weighted-graph estimator.

When using Nadaraya-Watson estimators, we chose and applied the kernel that gave the best results on average over the wind farm, and over the data records. We believe our method could benefit from replacing these estimators with an algorithm that would rather compute optimal weights efficiently and adaptively for each data point when performing the weighted $k$-NN imputation. By each data point we mean by wind turbine, and at each time step $t$ in the case of online weighted graphs. This may bring more robustness to the unweighted-graph imputation, since the method would adapt to any wind turbine without having to make any kernel assumption. The same applies to weighted-graph imputation, although some of the uncertainty is already handled through the addition of data information. Of course, the price to pay would be more computational effort.

A better imputation of missing values makes it easier to learn better forecasters, but imputation methods often require assumptions on distributions and some may be difficult to apply to any sort of predictor. Nearest-Neighbor imputation can be used with any predictor and do not ask for any other assumption than similar neighbors. In a highly nonstationary setup, such as offshore wind energy, the data point from the neighboring time series might just be one of the best estimates we can get for the data point we are missing.

\section*{Acknowledgments}
The authors gratefully acknowledge {\O}rsted for providing the data for the Westermost Rough offshore wind farm. The research leading to this work was carried out as a part of the Smart4RES project (European Union’s Horizon 2020, No. 864337). The sole responsibility of this publication lies with the authors. The European Union is not responsible for any use that may be made of the information contained therein.

\section*{Conflict of interest statement}
The authors report no conflict of interests.

\bibliography{windenergy}

\appendix

\section*{Appendix}

\nomenclature{\(\textbf{A}\)}{Adjacency matrix}
\nomenclature{\(\textbf{D}\)}{Degree matrix}
\nomenclature{\(E\)}{Set of edges}
\nomenclature{\(\textbf{f}\)}{Eigenvector}
\nomenclature{\(G\)}{Graph}
\nomenclature{\(K\)}{Kernel}
\nomenclature{\(x^{(i)}\)}{$x$ (e.g. power generation) of the $i$-th wind turbine among a varying set}
\nomenclature{\(x^{i}\)}{$x$ of wind turbine $i$ among the total set $i=1,\dots,N$}
\nomenclature{\(\textbf{L}\)}{Laplacian matrix}
\nomenclature{\(n_t\)}{Number of records available at time $t$}
\nomenclature{\(N\)}{Total number of wind turbines}
\nomenclature{\(s_t(i,j)\)}{Similarity between records of wind turbine $i$ and wind turbine $j$ at time $t$}
\nomenclature{\(l_t(i,j)\)}{Loss function we pay at time $t$ when guessing the similarity between records of wind turbine $i$ and wind turbine $j$}
\nomenclature{\(x_t^i\)}{$x$ of wind turbine $i$ at time $t$}
\nomenclature{\(t\)}{Time step}
\nomenclature{\(T\)}{Total number of records}
\nomenclature{\(V\)}{Set of nodes (or vertices)}
\nomenclature{\(v_i\)}{Node $i$}
\nomenclature{\(w^{(ji)}\)}{Weight of the $j$-th available record when imputing the $i$-th missing record with $k$-NN}
\nomenclature{\(\lambda\)}{Eigenvalue}
\nomenclature{\(r\)}{Dimension of the embeddings}
\nomenclature{\(\textbf{z}^i\)}{Embedding of wind turbine $i$}
\nomenclature{\(\eta\)}{Learning rate of lazy OGD}

\printnomenclature

\end{document}